\documentclass[12pt,a4paper]{iopart}

\usepackage{graphicx}
\usepackage{datetime}
\usepackage{color}
\usepackage{mathrsfs} 
\usepackage{dsfont}
\usepackage{xspace}
\usepackage{iopams}  
\usepackage{algorithm} 
\usepackage{algpseudocode}

\usepackage[bookmarks=true,
            bookmarksopen=true,colorlinks=false]{hyperref}

\newcommand{\eqref}{\eref}
\newcommand{\dd}{\mathrm{d}}
\newcommand{\ee}{\mathrm{e}}
\newcommand{\ii}{\mathrm{i}}
\newcommand{\R}{\mathds R}

\newcommand{\eps}{\varepsilon}

\newcommand{\E}{\mathcal E}

\newcommand{\JH}[1]{{\color{red}#1}}
\renewcommand{\JH}[1]{#1}
\newcommand{\Al}{\mathcal A_1}
\newcommand{\Ar}{\mathcal A_2}
\newcommand{\Hp}{{\mathcal H_\mathrm{p}}}
\newcommand{\Hf}{{\mathcal H_\mathrm{f}}}
\newcommand{\p}{_\mathrm{p}}

\renewcommand{\Re}{\mathrm{Re}}
\renewcommand{\Im}{\mathrm{Im}}

\def\Xint#1{\mathchoice
   {\XXint\displaystyle\textstyle{#1}}%
   {\XXint\textstyle\scriptstyle{#1}}%
   {\XXint\scriptstyle\scriptscriptstyle{#1}}%
   {\XXint\scriptscriptstyle\scriptscriptstyle{#1}}%
   \!\int}
\def\XXint#1#2#3{{\setbox0=\hbox{$#1{#2#3}{\int}$}
     \vcenter{\hbox{$#2#3$}}\kern-.5\wd0}}

\def\dashint{\Xint-}

\begin{document}

\title
[SGGTN solutions with polynomial initial data]
{Smooth Gowdy-symmetric generalised Taub--NUT solutions with polynomial initial data}

\author{J\"org Hennig}
\address{Department of Mathematics and Statistics,
           University of Otago,
           PO Box 56, Dunedin 9054, New Zealand}
\eads{\mailto{joerg.hennig@otago.ac.nz}}

\begin{abstract}
We consider smooth Gowdy-symmetric generalised Taub--NUT solutions, a class of inhomogeneous cosmological models with spatial three-sphere topology. They are characterised by existence of a smooth past Cauchy horizon and, with the exception of certain singular cases, they also develop a regular future Cauchy horizon. Several examples of exact solutions \JH{were previously} constructed, where the initial data (in form of the initial Ernst potentials) are polynomials of low degree. Here, we generalise to polynomial initial data of arbitrary degree. Utilising methods from soliton theory, we obtain a simple algorithm that allows us to construct the resulting Ernst potential with purely algebraic calculations. We also derive an explicit formula in terms of determinants, and we illustrate the method with two examples.
\\[2ex]{}
\end{abstract}

\section{Introduction\label{sec:intro}}

Some cosmological solutions to Einstein's field equations show the interesting behaviour of violating determinism and causality. Probably the best-known example is the Taub--NUT solution \cite{Taub1951, NUT1963, MisnerTaub1969}. 
The maximal globally hyperbolic development of this solution is the region between a past and a future Cauchy horizon. Beyond these, one can construct several non-equivalent extensions of the solution --- in contradiction to determinism. Also, in clear violation of causality, there are closed timelike curves beyond the Cauchy horizons.

In order to better understand properties of such peculiar solutions, which is particularly relevant in the context of the strong cosmic censorship conjecture \cite{Penrose1969, MoncriefEardley1981, Chrusciel1991} (see also \cite{Rendall2005, Ringstrom2009, Isenberg2015}), entire classes of solutions displaying a similar behaviour have been studied.

The large class of the \emph{generalised Taub--NUT spacetimes} was introduced by Moncrief \cite{Moncrief1984}, which, in contrast to the very symmetric, spatially \emph{homogeneous} Taub--NUT solution, describe \emph{inhomogeneous} cosmological models. These solutions have a $U(1)$ isometry group, and existence was shown under the assumption of analyticity.

The focus of this paper will be another class of inhomogeneous models with spatial three-sphere topology $\mathbb S^3$, the \emph{smooth Gowdy-symmetric generalised Taub--NUT (SGGTN) solutions}, which was first introduced in \cite{BeyerHennig2012}. The SGGTN solutions are smooth solutions, and they have two spacelike Killing vectors (Gowdy symmetry). They are characterised by existence of a regular past Cauchy horizon, and  (with the exception of certain singular cases, which can already be identified from the data at the past horizon) they develop a second horizon, a regular future Cauchy horizon. Extensions beyond both horizons can be constructed, and we again observe a breakdown of determinism and causality.

The initial local and global existence considerations for SGGTN solutions in \cite{BeyerHennig2012} assumed that the past horizon has closed null generators. Later, the class of SGGTN solutions was extended to also include past horizons with \emph{non-closed} generators \cite{Hennig2016b}. The study of this type of horizon is relevant, since some investigations of cosmological Cauchy horizons assume closedness of the generators (see, e.g., \cite{FriedrichRaczWald1999,MoncriefIsenberg1983,Racz2000}). Hence, it is important to study what happens if this assumption is violated. On the other hand, horizons with closed generators seem to be more generic in the sense that they can occur in less symmetric spacetimes \cite{MoncriefIsenberg2018,PetersenRacz2018}. Another generalisation of the SGGTN solutions was introduced in \cite{Hennig2019}, where an additional electromagnetic field was considered and solutions to the Einstein--Maxwell equations in \emph{electrovacuum} were studied. Also in the electromagnetic case, past and future Cauchy horizons are generally present, and interesting singular cases occur if the initial data violate a regularity condition. Here, however, we will consider SGGTN solutions in \emph{vacuum}.

The basis for the investigations of global properties of SGGTN solutions in \cite{BeyerHennig2012} and \cite{Hennig2016b} was the remarkable fact that the essential part of the Einstein vacuum equations for Gowdy-symmetric spacetimes can be reformulated in terms of the \emph{Ernst equation} --- a single, complex equation, which belongs to the class of soliton/integrable equations.
This enables abstract considerations with soliton methods, and related techniques can even be used to construct examples of exact solutions. Using ``Sibgatullin's integral method'', three families of exact solutions with cubic initial Ernst potentials at the past Cauchy horizon have been constructed in \cite{BeyerHennig2014, Hennig2016b, Hennig2016a}. 

Here, we extend these considerations and study an initial value problem for the Ernst equation where the initial Ernst potential is a polynomial of \emph{arbitrary} degree (in a coordinate $x$ to be introduced below). After carrying out all required quadratures, we formulate a simple algorithm that enables us to obtain the Ernst potential in the entire ``Gowdy square'', i.e.\ the region between the Cauchy horizons, from the initial data at the past horizon. The algorithm only requires algebraic operations, including solving a linear system of equations. Alternatively, we can also represent the final Ernst potential as a quotient of determinants.  

We begin our considerations in Sec.~\ref{sec:metric} by introducing suitable coordinates for SGGTN solutions, and we obtain the field equations and explain their reformulation in terms of the Ernst equation. Afterwards, in Sec.~\ref{sec:IVP}, we solve an initial value problem for the Ernst equation with polynomial initial data. For that purpose, we first summarise Sibgatullin's integral method. Then we describe how to choose polynomial initial data subject to conditions required in the context of SGGTN solutions. Afterwards, we reformulate the problem in terms of algebraic equations and describe the algorithm to obtain the Ernst potential everywhere. Next, we also derive formulae for two of the metric potentials, which are closely related to the Ernst potential. Finally, in Secs.~\ref{sec:Ex1} and \ref{sec:Ex2}, we apply the method to initial Ernst potentials of fourth and sixth degrees, respectively.

\section{Metric and field equations}\label{sec:metric}

Initially, we express SGGTN spacetimes in the coordinates $(t,\theta,\tilde\rho_1,\tilde\rho_2)$ used in \cite{BeyerHennig2012}\footnote{In \cite{BeyerHennig2012}, these coordinates were labelled $(t,\theta,\rho_1,\rho_2)$. The additional tildes in our present notation will allow us to distinguish them from the coordinates that are adapted to the generators of the past horizon, which we introduce below.}
with line element
\begin{equation}\label{eq:metric0}
  \dd s^2=\ee^M(-\dd t^2+\dd\theta^2)
  +R_0\left[\sin^2\!t\,\ee^{\tilde u} (\dd\tilde\rho_1+\tilde Q\, \dd\tilde\rho_2)^2+\sin^2\!\theta\,\ee^{-\tilde u}\, \dd\tilde\rho_2^2\right].
\end{equation}
Here, $R_0>0$ is a constant, and $\tilde u$, $\tilde Q$ and $M$ are functions of $t$ and $\theta$ alone. The two Killing vectors for Gowdy symmetry are the  coordinate vectors $\tilde\xi=\partial_{\tilde\rho_1}$, $\tilde\eta=\partial_{\tilde\rho_2}$.
Moreover, the angles $\tilde\rho_1$, $\tilde\rho_2$ are defined in the regions
\begin{equation}\label{eq:rhodomain}
 \frac{\tilde\rho_1+\tilde\rho_2}{2}\in(0,2\pi),\quad
 \frac{\tilde\rho_1-\tilde\rho_2}{2}\in(0,2\pi).
\end{equation}
\JH{The} existence of solutions is initially guaranteed in the ``Gowdy square''
\begin{equation}
 t\in(0,\pi),\quad \theta\in(0,\pi),
\end{equation}
where the boundaries $\theta=0,\pi$ correspond to the symmetry axes $\Al$ and $\Ar$, and the past and future Cauchy horizons $\Hp$ and $\Hf$ are located at $t=0$ and $t=\pi$, respectively. 

In the original local and global existence proofs in \cite{BeyerHennig2012}, the SGGTN solutions were subject to the restriction that the past horizon $\Hp$ is generated by the Killing vector $\tilde\xi$ alone. This corresponds to a horizon with closed null generators. In \cite{Hennig2016b}, this was extended to Cauchy horizons with non-closed null generators, where $\Hp$ is generated by a more general linear combination $\tilde\xi-\tilde a\p\tilde\eta$, $\tilde a\p=\mathrm{constant}\neq \pm1$ (see below). However, this case can almost be reduced to the previous one: The coordinate transformation $(\tilde\rho_1,\tilde\rho_2)\to(\rho_1,\rho_2)$ with
\begin{equation}\label{eq:trafo}
 \tilde\rho_1=\rho_1-\tilde a\p\rho_2,\quad
 \tilde\rho_2=-\tilde a\p\rho_1+\rho_2
\end{equation}
and a corresponding change of the Killing basis to the new vectors $\xi=\partial_{\rho_1}$, $\eta=\partial_{\rho_2}$ achieves that the more general type of horizon is generated by $\xi$ alone. The only differences are that the metric functions $\tilde u$ and $\tilde Q$ transform into new functions $u$ and $Q$, that $M$ satisfies slightly modified boundary conditions, and that the coordinates $(\rho_1, \rho_2)$ are defined in a different domain than $(\tilde\rho_1, \tilde\rho_2)$, which is obtained by transforming \eqref{eq:rhodomain} with \eqref{eq:trafo}, see \cite{Hennig2016b} and Appendix A in \cite{Hennig2016a} for more details.

As shown in \cite{Hennig2016b}, the metric potentials must satisfy the following boundary conditions at the axes $\Al$ ($\theta=0$) and $\Ar$ ($\theta=\pi$),
\begin{equation}\label{eq:regcond1}\fl
 \Al: \quad Q=1,\quad \ee^{M+u}=\frac{R_0}{(1+\tilde a\p)^2},\qquad
 \Ar: \quad Q=-1,\quad \ee^{M+u}=\frac{R_0}{(1-\tilde a\p)^2}.
\end{equation}
Information about the generator of the past horizon in terms of the original coordinates enters these conditions in form of the constant $\tilde a\p$. Note that the conditions in \eqref{eq:regcond1} become singular for $\tilde a\p=\pm1$. However, as discussed in \cite{Hennig2016b}, these values for $\tilde a\p$ are not permitted, since the past horizon $\Hp$ would then contain a singularity. Since the requirement of a regular past Cauchy horizon $\Hp$ is part of the definition of SGGTN solutions, this is not allowed.

Finally, in order to obtain a more convenient representation without trigonometric functions, we replace the coordinates $t$ and $\theta$ by
\begin{equation}
 x=\cos\theta,\quad y=\cos t,
\end{equation}
in terms of which the line element reads
\begin{equation}\fl\label{eq:metric}
   \dd s^2 =\ee^M\left(\frac{\dd x^2}{1-x^2}-\frac{\dd y^2}{1-y^2}\right)+R_0\left[(1-y^2)\,\ee^u 
   (\dd\rho_1+Q\,\dd\rho_2)^2+(1-x^2)\,\ee^{-u}\, \dd\rho_2^2\right].
\end{equation}
The Einstein vacuum equations\footnote{The Einstein equations in terms of the coordinates $(t,\theta,\rho_1,\rho_2)$ can be found in \cite{BeyerHennig2012}. Here, we give the equations in terms of $(x,y,\rho_1,\rho_2)$, which can be obtained from the more general Einstein--Maxwell equations for electrovacuum Gowdy spacetimes in \cite{Hennig2016a},  \cite{Hennig2019} in the limit of vanishing electromagnetic field.} for this metric consist of two second-order equations for $u$ and $Q$,
\begin{eqnarray}\label{eq:EinsteinEQu}
  \fl
  0 &=& (1-x^2)u_{,xx}-(1-y^2)u_{,yy} -\frac{1-y^2}{1-x^2}\,\ee^{2u}
        \left[(1-x^2)Q_{,x}^{\ 2}-(1-y^2)Q_{,y}^{\ 2}\right]\nonumber\\
  \fl      
	 && -2xu_{,x}+2yu_{,y}+2,
\end{eqnarray}
\begin{equation}\label{eq:EinsteinEQQ}\fl
    0=(1-x^2)Q_{,xx}-(1-y^2)Q_{,yy}+2(1-x^2)Q_{,x}u_{,x}
    -2(1-y^2)Q_{,y}u_{,y}+4yQ_{,y}
\end{equation}    
and two first-order equations for $M$,
\begin{eqnarray}\label{eq:Mx}
  \fl
   M_{,x} &=& -\frac{1-y^2}{2(x^2-y^2)}\Big[
    x(1-x^2)u_{,x}^{\ 2}+x(1-y^2)u_{,y}^{\ 2}-2y(1-x^2)u_{,x}u_{,y}
     \nonumber\\
  \fl   
   &&
   +2\frac{x^2+y^2-2x^2y^2}{1-y^2}u_{,x}-4xyu_{,y}-4x\nonumber\\
  \fl 
   &&
   +\frac{1-y^2}{1-x^2}\ee^{2u}
   \Big(x(1-x^2)Q_{,x}^{\ 2}+ x(1-y^2)Q_{,y}^{\ 2}
   -2y(1-x^2)Q_{,x}Q_{,y}\Big)\Big],
  \label{eq:Meq1}
  \end{eqnarray}
\begin{eqnarray}\label{eq:My}
  \fl
   M_{,y} &=& \frac{1-x^2}{2(x^2-y^2)}\Big[
    y(1-x^2)u_{,x}^{\ 2}+y(1-y^2)u_{,y}^{\ 2}-2x(1-y^2)u_{,x}u_{,y}\nonumber\\
  \fl  
   &&
   +4xy u_{,x}-2\frac{x^2+y^2-2x^2y^2}{1-x^2}u_{,y}-4y\nonumber\\
   \fl 
   &&
   +\frac{1-y^2}{1-x^2}\ee^{2u}
   \Big(y(1-x^2)Q_{,x}^{\ 2}+ y(1-y^2)Q_{,y}^{\ 2}
   -2x(1-y^2)Q_{,x}Q_{,y}\Big)\Big].\label{eq:Meq2}
  \end{eqnarray}
Since $M$ does not appear in \eqref{eq:EinsteinEQu}, \eqref{eq:EinsteinEQQ}, these equations can be solved first. Afterwards, $M$ can be obtained from \eqref{eq:Meq1}, \eqref{eq:Meq2} via line integration.

\JH{In the following, we describe how \eqref{eq:EinsteinEQu} and \eqref{eq:EinsteinEQQ} can be reformulated into the Ernst equation.}
For that purpose, we define functions $f$ and $a$ in terms of the Killing vectors,
\begin{eqnarray}
  \label{eq:f}
  f &=& \frac{1}{R_0}g(\eta,\eta)
        \equiv Q^2\ee^u(1-y^2)+\ee^{-u}(1-x^2),\\
  \label{eq:a}
  a &=& \frac{g(\xi,\eta)}
             {g(\eta,\eta)}
        \equiv \frac{Q}{f}\ee^u(1-y^2),
\end{eqnarray}
\JH{and rewrite \eqref{eq:EinsteinEQu}, \eqref{eq:EinsteinEQQ} as a system for $f$ and $a$. Taking suitable linear combinations of the resulting equations, we obtain the following second-order equations for $f$ and $a$,
\begin{eqnarray}\label{eq:EinsteinEQf}
 f\left[(1-x^2)f_{,xx} - 2x f_{,x} - (1-y^2)f_{,yy} + 2yf_{,y}\right]
 \nonumber\\
 \qquad 
 = (1-x^2)\left[f_{,x}^{\ 2} - \frac{f^4 a_{,y}^{\ 2}}{(1-x^2)^2}\right]
 + (1-y^2)\left[f_{,y}^{\ 2} - \frac{f^4 a_{,x}^{\ 2}}{(1-y^2)^2}\right],\\
 \left(\frac{f^2 a_{,x}}{1-y^2}\right)_{,x} 
  =\left(\frac{f^2 a_{,y}}{1-x^2}\right)_{,y}.
\end{eqnarray}
The latter equation can easily be solved by introducing a potential $b$ satisfying
\begin{equation}
 b_{,x} = \frac{f^2 a_{,y}}{1-x^2},\quad
 b_{,y} = \frac{f^2 a_{,x}}{1-y^2}.
\end{equation}
Conversely, once $b$ is known, we can reconstruct $a$ via line integration from 
\begin{equation}\label{eq:a1}
  a_{,x} = \frac{1-y^2}{f^2}b_{,y},\quad
  a_{,y} = \frac{1-x^2}{f^2}b_{,x}.
\end{equation}
This, however, only leads to a well-defined function $a$ if the integrability condition $a_{,xy}=a_{,yx}$ is satisfied. Using \eqref{eq:a1}, this translates into the equation
\begin{equation}\label{eq:ErnstIm}
\fl
 f\cdot\left[(1-x^2)b_{,xx} - 2x b_{,x} - (1-y^2)b_{,yy} + 2yb_{,y}\right]
 =2(1-x^2)f_{,x} b_{,x}-2(1-y^2)f_{,y} b_{,y}
\end{equation}
for $b$. Moreover, replacing $a$ in favour of $b$ in \eqref{eq:EinsteinEQf}, we obtain
\begin{eqnarray}\label{eq:ErnstRe}
 f\cdot\left[(1-x^2)f_{,xx} - 2x f_{,x} - (1-y^2)f_{,yy} + 2yf_{,y}\right]
 \nonumber\\
 \qquad =(1-x^2)\left(f_{,x}^{\ 2} - b_{,x}^{\ 2}\right)
 + (1-y^2)\left(f_{,y}^{\ 2} - b_{,y}^{\ 2}\right).
\end{eqnarray}
The two equations \eqref{eq:ErnstRe} and \eqref{eq:ErnstIm} can be taken to be the real and imaginary parts of a single complex equation. For that purpose, we finally define the Ernst potential
\begin{equation}\label{eq:E}
  \E   = f+\ii b,
\end{equation}
which, as a consequence of \eqref{eq:ErnstRe}
and \eqref{eq:ErnstIm}, 
needs to satisfy the \emph{Ernst equation}
\begin{equation}\label{eq:ErnstEQ}
\fl
   f\cdot\left[(1-x^2)\E_{,xx}-2x\E_{,x}-(1-y^2)\E_{,yy}+2y\E_{,y}\right]
   = (1-x^2)\E_{,x}^{\ 2}
       -(1-y^2)\E_{,y}^{\ 2}.
\end{equation}
}

The Ernst equation was originally derived to describe axisymmetric and stationary vacuum spacetimes (with one spacelike and one timelike Killing vector) \cite{Ernst1968, KramerNeugebauer1968}. 
\JH{However, the corresponding cylindrical Weyl--Lewis--Papapetrou coordinates $(\rho,\zeta,\varphi,t)$ are directly related to our coordinates for Gowdy spacetimes (with two spacelike Killing vectors) through the formal, complex coordinate transformation
\begin{equation}\label{eq:transformation}
 \rho=\ii\sin t\sin\theta 
     \equiv \ii\sqrt{(1-x^2)(1-y^2)},\quad
 \zeta=\cos t\cos\theta
      \equiv x y.
\end{equation}
}
In particular, in either setting, the \emph{nonlinear} Ernst equation belongs to the remarkable class of integrable equations, which can be reformulated as an integrability condition of a \emph{linear} matrix problem. Some related applications for axisymmetric and stationary spacetimes can be found in
\cite{Neugebauer2003, NeugebauerHennig2012}, and for SGGTN solutions, soliton methods were crucial in the analysis of global existence \cite{BeyerHennig2012} and for the construction of exact solutions 
\cite{BeyerHennig2014, Hennig2016b, Hennig2016a}. Note that the Ernst equation and the underlying linear problem also generalise to SGGTN solutions in electrovacuum \cite{Hennig2016a,Hennig2019}.

The Ernst formulation is also particularly useful for determining the regularity of the solution from the initial data. If we denote the values of $b=\Im(\E)$ at the points $\theta=0$, $t=0$ (i.e.\ $x=y=1$) and $\theta=\pi$, $t=0$ (i.e.\ $x=-1$, $y=1$) on the past horizon $\Hp$ by $b_A$ and $b_B$, respectively, then the SGGTN solutions will be regular up to $t=\pi$ ($y=-1$) and have a regular future Cauchy horizon there whenever the initial data satisfy the regularity condition
\cite{BeyerHennig2012, Hennig2016b}
\begin{equation}
 \label{eq:regularity}
 b_B-b_A\neq \pm 4.
\end{equation}
On the other hand, for initial data with $b_B-b_A=4$, the solution develops a curvature singularity at $\theta=0$, $t=\pi$, and if $b_B-b_A=-4$ holds, then a singularity is located at $\theta=\pi$, $t=\pi$. We construct new examples of regular and singular solutions in Secs.~\ref{sec:Ex1} and \ref{sec:Ex2} below.

\section{Initial value problem with polynomial Ernst potential}
\label{sec:IVP}

\subsection{Sibgatullin's integral method}

We intend to solve the Ernst equation for a polynomial initial Ernst potential at $t=0$ (i.e.\ $y=1$). According to \eqref{eq:transformation}, this is equivalent to the construction of an Ernst potential for an axisymmetric and stationary spacetime with data on the symmetry axis $\rho=0$ in the interval $\zeta\in[-1,1]$. This problem can be elegantly solved with ``Sibgatullin's integral method'' \cite{Sibgatullin1984, MankoSibgatullin1993}, where the initial value problem is reformulated in terms of a linear integral equation. This, in turn, can be further reduced to a system of linear (algebraic) equations. 

We start with the original formulation for axisymmetric and stationary spacetimes with cylindrical coordinates $\rho$, $\zeta$, but we can  translate at any time into our coordinates using \eqref{eq:transformation}.
The main ingredient is the following linear integral equation for a complex function $\mu(\xi;\rho,\zeta)$ 
\JH{[cf.~Eq.~(2.44) in \cite{MankoSibgatullin1993}, in the special case of vanishing electromagnetic field $f(\xi)$],}
\begin{equation}\label{eq:inteq}
 \dashint_{-1}^1\frac{\mu(\xi;\rho,\zeta)[e(\xi)+\tilde e(\eta)]\,\dd\sigma}{(\sigma-\tau)\sqrt{1-\sigma^2}}=0.
\end{equation}
Here, $\dashint$ denotes the principal value integral, and we have defined 
$\xi:=\zeta+\ii\rho\sigma$ and $\eta:=\zeta+\ii\rho\tau$ with $\sigma,\tau\in[-1,1]$. 
Moreover, the functions
\begin{equation}
 e(\xi):=\E(\rho=0,\zeta=\xi)
 \quad\mbox{and}\quad
 \tilde e(\xi):=\overline{e(\bar\xi)},
\end{equation}
where the bar denotes complex conjugation,
are obtained from the analytic continuation of the axis Ernst potential at $\rho=0$.
A unique solution to \eqref{eq:inteq} is fixed by imposing the additional constraint
\begin{equation}\label{eq:constraint}
 \int_{-1}^1\frac{\mu(\xi;\rho,\zeta)\,\dd\sigma}{\sqrt{1-\sigma^2}}=\pi.
\end{equation}
The resulting Ernst potential is then given by
\begin{equation}\label{eq:EP}
 \E(\rho,\zeta)=\frac{1}{\pi}\int_{-1}^1\frac{e(\xi)\mu(\xi)\dd\sigma}{\sqrt{1-\sigma^2}},
\end{equation}
 where $\mu(\xi)$ is an abbreviation for $\mu(\xi;\rho,\zeta)$.

From the Ernst potential, we can construct the auxiliary quantity $a$ by solving \eqref{eq:a1}. However, much more conveniently, $a$ can also be calculated directly from $\mu$,
\begin{equation}\label{eq:a2}
 a=\frac{2}{\pi f}\,\Im\int_{-1}^1\frac{\xi\mu(\xi)\,\dd\sigma}{\sqrt{1-\sigma^2}}.
\end{equation}
The metric potentials $u$ and $Q$ can then be obtained algebraically from \eqref{eq:f}, \eqref{eq:a},
\begin{equation}\label{eq:uQ}
 \ee^u=\frac{f a^2}{1-y^2}+\frac{1-x^2}{f},\quad
 Q=\frac{f^2 a}{f^2 a^2 + (1-x^2) (1-y^2)}.
\end{equation}
Only the calculation of the remaining metric function $M$ does require integration to solve \eqref{eq:Meq1}, \eqref{eq:Meq2}. Here, however, we will restrict our attention to the Ernst potential $\E$ and the metric functions $u$ and $Q$.
\subsection{Initial data}

We prescribe the initial Ernst potential at the past horizon $\Hp$, i.e.\ at $t=0$ ($y=1$). According to \eqref{eq:transformation}, we have $\zeta=x$ at $y=1$. Hence, in our analogy between axisymmetric spacetimes and SGGTN solutions, we can interchangeably use $\zeta$ or $x$ as argument for the initial data. We assume that the initial Ernst potential at $\Hp$, which we denote by $\E\p=f\p+\ii b\p$, is a \emph{polynomial of degree} $N$ in $\zeta$.
Note that polynomial data would not be an appropriate choice in the axisymmetric and stationary setting, where the Ernst potential in an asymptotically flat spacetime should satisfy $\E\to 1$ as $\zeta\to\pm\infty$, but are perfectly fine in the Gowdy context with $\zeta\in[-1,1]$.

Specifically, we take $f\p$ and $b\p$ to have the form
\begin{equation}\label{eq:initial_data}
 f\p(\zeta)=c\prod_{m=1}^N(\zeta-\xi_m),\quad
 b\p(\zeta)=\sum_{k=0}^Nd_k\zeta^k.
\end{equation}
The imaginary part $b\p$ is a general real polynomial with coefficients $d_k\in\R$, $k=0,\dots,N$, whereas we choose the real part $f\p$ to be given in terms of a constant $c$ and \emph{real and distinct} zeros $\xi_m\in\R$, $m=1,\dots N$. Furthermore, both polynomials are subject to the following constraints \cite{BeyerHennig2014},
\begin{equation} \label{eq:Ernstconstraints}
 \fl
 f\p(\zeta=\pm1)=0,\quad
 f\p(\zeta)>0\quad\mbox{for}\quad \zeta\in(-1,1),\quad
 \frac{\dd b\p}{\dd\zeta}(\zeta=\pm1)=\mp 2.
\end{equation}
Consequently, two of the zeros of $f\p$ need to be equal to $1$ and $-1$, while all other roots must satisfy $|\xi_m|>1$. Moreover, the sign of $c$ is fixed by the second condition in \eqref{eq:Ernstconstraints}, and the polynomial coefficients $d_k$ must be chosen in accordance with the third condition. 

As we will see in an example below (see Sec.~\ref{sec:Ex1}), we can relax the requirement of distinct zeros of $f\p$ and also choose initial data with \emph{repeated} roots, by obtaining these as a limit of a sequence of initial functions with distinct roots. Furthermore, the procedure also works if some (or all) 	zeros are pairs of complex conjugate roots. However, to avoid more involved and unpleasant discussions of complex square roots, we will always assume in the following presentation that all $\xi_m$ are real.

\subsection{Algebraic equations}

Following \cite{MankoSibgatullin1993}, we convert the integral equation \eqref{eq:inteq} into a system of linear equations by first finding the roots of the algebraic equation
\begin{equation}
 0=e(\xi)+\tilde e(\xi)\equiv 2f\p(\xi).
\end{equation}
Hence, the required roots are exactly the zeros $\xi_m$ of $f\p$, which explains why we have chosen a product representation of $f\p$ in \eqref{eq:initial_data}. The solution $\mu$ to the integral equation (with our assumption of single roots $\xi_m$) should then be of the form \cite{MankoSibgatullin1993}
\begin{equation}\label{eq:mu}
 \mu(\xi;\rho,\zeta)=A_0(\rho,\zeta)+\sum_{m=1}^N\frac{A_m(\rho,\zeta)}{\xi-\xi_m}.
\end{equation}
This already fixes the $\xi$-dependence of $\mu$, and it remains to find the $(\rho,\zeta)$-dependence in the form of the unknown functions $A_m(\rho,\zeta)$, $m=0,\dots,N$. We will see that plugging \eqref{eq:mu} into the integral equation \eqref{eq:inteq} gives the condition that a polynomial of degree $N-1$ in $\eta$ must vanish, which will only be the case if the $N$ polynomial coefficients vanish separately. Supplemented by one further condition from the constraint \eqref{eq:constraint}, we obtain an algebraic system of $N+1$ equations for the $N+1$ unknowns $A_0$, $\dots$, $A_N$.

With 
$e(\xi)=f\p(\xi)+\ii b\p(\xi)$ 
and 
$\tilde e(\eta)=f\p(\eta)-\ii b\p(\eta)$,
and using
$\sigma-\tau=(\xi-\eta)/(\ii\rho)$, the integrand in \eqref{eq:inteq} (without the square root factor) becomes
\begin{equation}\label{eq:integrand}
\fl
 \frac{\mu(\xi)[e(\xi)+\tilde e(\eta)]}{\sigma-\tau} 
 = \ii\rho\frac{f\p(\xi)+\ii[b\p(\xi)-b\p(\eta)]+f\p(\eta)}{\xi-\eta}
   \left(A_0+\sum_{m=1}^N\frac{A_m}{\xi-\xi_m}\right).
\end{equation}
The required $\sigma$-integration (recall that $\sigma$ is contained in $\xi=\zeta+\ii\rho\sigma$) can be done via a partial fraction decomposition. Thereby, in order to systematically obtain a reasonably concise result, it is useful to separately consider the decompositions for each of the three terms $f\p(\xi)$, $b\p(\xi)-b\p(\eta)$ and $f\p(\eta)$. Since $\xi-\xi_m$ is a factor of $f\p(\xi)$, this term is conveniently first divided by this factor. On the other hand, $\xi-\eta$ is a factor of $b\p(\xi)-b\p(\eta)$, so we can first split off this factor from the second term. Finally, the third term $f\p(\eta)$ is divisible by $\eta-\xi_m$, which is useful, since $1/(\eta-\xi_m)$ appears in the decomposition
\begin{equation}\label{eq:decomp}
 \frac{1}{(\xi-\xi_m)(\xi-\eta)}=\frac{1}{\eta-\xi_m}
  \left(\frac{1}{\xi-\eta}-\frac{1}{\xi-\xi_m}\right).
\end{equation}

We start with the partial fraction decomposition of the first term containing $f\p(\xi)$. For that purpose, in addition to the product representation in \eqref{eq:initial_data}, we also introduce the power form
\begin{equation}\label{eq:ck}
 f\p(\xi)
 \equiv c\prod_{m=1}^N(\xi-\xi_m)
 =\sum_{k=0}^N c_k\xi^k,
\end{equation}
where the polynomial coefficients $c_k$, $k=0,\dots,N$, can easily be found by expanding the product representation. We first divide $f\p(\xi)$ by $\xi-\xi_m$ and also introduce a power form for the resulting quotients,
\begin{equation}\label{eq:cmk}
 \frac{f\p(\xi)}{\xi-\xi_m}
 \equiv c\prod_{\begin{minipage}{8mm}\scriptsize$l=1$\\$l\neq m$\end{minipage}}^N(\xi-\xi_l)
 =\sum_{k=0}^{N-1}c_{mk}\xi^k.
\end{equation}
The coefficients $c_{mk}$ could be found by expanding the power form, but if we already have the coefficients $c_k$, they can easily be given explicitly as
\begin{equation}\label{eq:cmk1}
 \fl
 c_{mk} = \sum_{l=k+1}^N c_l\xi_m^{l-(k+1)}
 \equiv\sum_{l=0}^{N-k-1} c_{k+l+1}\xi_m^{l},
 \quad 
 m=1,\dots,N,\quad 
 k=0,\dots,N-1.
\end{equation}
Secondly, we divide by $\xi-\eta$. This gives, after a short calculation\footnote{Note that the comma in $c_{m, k+l+1}$ only separates the two indices, i.e.\ it does not represent any form of differentiation.},
\begin{eqnarray}\label{eq:term1}
\fl\nonumber
 \ii\rho\frac{f\p(\xi)\mu(\xi)}{\xi-\eta}\\
 \fl
 \qquad= \ii\rho\sum_{l=0}^{N-1}
    \left[
      A_0\sum_{k=0}^{N-l-1} c_{k+l+1}\xi^k
      +\sum_{m=1}^N\left(\sum_{k=0}^{N-l-2} c_{m, k+l+1}\xi^k\right)A_m\right]
    \eta^l+\frac{R_1(\rho,\zeta,\eta)}{\xi-\eta}.
\end{eqnarray}
The function $R_1$ in the last term summarises a few additional contributions that depend on $\xi$ only in the form $1/(\xi-\eta)$. As we will see below [cf.~\eqref{eq:int1}], all such terms are irrelevant as they integrate to zero.

Next, we consider the second term in \eqref{eq:integrand}, i.e.\ the term proportional to $b\p(\xi)-b\p(\eta)$. We start with dividing by $\xi-\eta$,
\begin{equation}
 \frac{b\p(\xi)-b\p(\eta)}{\xi-\eta}
 = \sum_{k=0}^N d_k\frac{\xi^k-\eta^k}{\xi-\eta}
 = \sum_{k=0}^N\sum_{l=0}^{k-1} d_k\xi^l\eta^{k-1-l}.
\end{equation}
Afterwards, we perform the divisions by $\xi-\xi_m$. The result is (after a somewhat more lengthy calculation than above)
\begin{eqnarray}\label{eq:term2}
\fl\nonumber
 \ii\rho\frac{[b\p(\xi)-b\p(\eta)]\mu(\xi)}{\xi-\eta}\\
 \fl
 \qquad= -\rho\sum_{l=0}^{N-1}
  \left[
    A_0\sum_{k=0}^{N-l-1} d_{k+l+1}\xi^k
    + \sum_{m=1}^N A_m\left(\sum_{k=0}^{N-l-2}e_{m,k+l+1}\xi^k+\frac{e_{ml}}{\xi-\xi_m}\right)
  \right]\eta^l,
\end{eqnarray}
where we have defined the constants
\begin{equation}\label{eq:eml}
 e_{ml} = \sum_{k=0}^{N-l-1} d_{l+k+1}\xi_m^k.
\end{equation}

Finally, we consider the third term in \eqref{eq:integrand}, which is proportional to $f\p(\eta)$. Using \eqref{eq:decomp} and the fact that $\eta-\xi_m$ is a factor of $f\p(\eta)$, we obtain 
\begin{equation}\label{eq:term3}
 \ii\rho\frac{f\p(\eta)\mu(\xi)}{\xi-\eta}
 = -\ii\rho\sum_{l=0}^{N-1}
   \left(\sum_{m=1}^N\frac{c_{ml}}{\xi-\xi_m}A_m\right)\eta^l
   +\frac{R_2(\rho,\zeta,\eta)}{\xi-\eta},
\end{equation}
where the last term again represents additional contributions that integrate to zero.

We can now add the decompositions \eqref{eq:term1}, \eqref{eq:term2} and \eqref{eq:term3} of the three terms in the integral equation \eqref{eq:inteq}. Notably, the constants $c_k$, $d_k$, and $c_{mk}$, $e_{mk}$ only appear in the complex combinations
\begin{equation}\label{eq:alphabeta}
 \alpha_k   := c_k+\ii d_k,\quad
 \beta_{mk} := c_{mk}+\ii e_{mk},
\end{equation}
where the constants $\alpha_k$ are simply the polynomial coefficients for the initial Ernst potential, $\E\p(\zeta)=\sum_{k=0}^N\alpha_k\zeta^k$.
If we also include the square root factor, then the full integrand of \eqref{eq:inteq} becomes
\begin{eqnarray}\label{eq:inteq_decomp}
 \fl\nonumber
 \frac{\mu(\xi;\rho,\zeta)[e(\xi)+\tilde e(\eta)]}{(\sigma-\tau)\sqrt{1-\sigma^2}}\\
 \fl\nonumber
 \quad=
  \frac{\ii\rho}{\sqrt{1-\sigma^2}}
  \sum_{l=0}^{N-1}
  \left[
   A_0\sum_{k=0}^{N-l-1}\alpha_{k+l+1}\xi^k
   +\sum_{m=1}^N A_m
    \left(\sum_{k=0}^{N-l-2}\beta_{m,k+l+1}\xi^k-\frac{\bar\beta_{ml}}{\xi-\xi_m}\right)
  \right]\eta^l\\
  \fl
  \qquad +\frac{R(\rho,\zeta,\eta)}{\sqrt{1-\sigma^2}(\sigma-\tau)},
  \quad
  R:=R_1+R_2.
\end{eqnarray}

Before we integrate this expression, we also consider the constraint \eqref{eq:constraint} and the formula for the Ernst potential \eqref{eq:EP}. For the constraint, no further decomposition is required, and we immediately have
\begin{equation}\label{eq:constraint_decomp}
 \frac{\mu(\xi;\rho,\zeta)}{\sqrt{1-\sigma^2}}
  = \frac{1}{\sqrt{1-\sigma^2}}
    \left(
      A_0+\sum_{m=1}^N\frac{A_m}{\xi-\xi_m}
	 \right).
\end{equation}
In case of the Ernst potential, a quick calculation gives the partial fraction decomposition
\begin{equation}\label{eq:Ernst_decomp}
\fl
 \frac{\mu(\xi;\rho,\zeta)e(\xi)}{\sqrt{1-\sigma^2}}
 = \frac{1}{\sqrt{1-\sigma^2}}
   \left[
    A_0\sum_{k=0}^N\alpha_k\xi^k
    +\sum_{m=1}^N A_m
      \left(\sum_{k=0}^{N-1}\beta_{mk}\xi^k+\frac{\ii b\p(\xi_m)}{\xi-\xi_m}\right)
   \right].
\end{equation}
Note that we could also replace $\ii b\p(\xi_m)$ with $\E\p(\xi_m)$, since $f\p(\xi_m)=0$.

Next, we integrate \eqref{eq:inteq_decomp}, \eqref{eq:constraint_decomp} and \eqref{eq:Ernst_decomp}, i.e.\ we plug the decomposed integrands into \eqref{eq:inteq}, \eqref{eq:constraint} and \eqref{eq:EP}. For that purpose, we calculate the following integrals,
\begin{equation}\label{eq:int1}
 \dashint_{-1}^1\frac{\dd\sigma}{\sqrt{1-\sigma^2}(\sigma-\tau)}=0,
\end{equation}
\begin{equation}\label{eq:int2}
 \fl
 p_n(\rho,\zeta):=\frac{1}{\pi}\int_{-1}^1\frac{\xi^n}{\sqrt{1-\sigma^2}}\,\dd\sigma
 = (\rho^2+\zeta^2)^{\frac{n}{2}}P_n\left(\frac{\zeta}{\sqrt{\rho^2+\zeta^2}}\right),
 \quad n=0,1,2,\dots,
\end{equation}
\begin{equation}\label{eq:int3}
\fl
 w_m(\rho,\zeta):=\frac{1}{\pi}\int_{-1}^1\frac{\dd\sigma}{\sqrt{1-\sigma^2}(\xi-\xi_m)}
 = \frac{\mathrm{sgn}(\zeta-\xi_m)}{\sqrt{\rho^2+(\zeta-\xi_m)^2}},
 \quad m=1,\dots,N,
\end{equation}
where $P_n$ denotes the $n$th Legendre polynomial.
The derivations of these formulae are sketched in \ref{sec:AppIntegrals}.
Note that the functions $p_n(\rho,\zeta)$, despite the square roots in \eqref{eq:int2}, simplify to polynomials in $\rho$ and $\zeta$ (of degree $2\lfloor n/2\rfloor$ in $\rho$ and degree $n$ in $\zeta$). In terms of the actually relevant coordinates $x\in[-1,1]$, $y\in[-1,1]$ for SGGTN solutions, $p_n$ and $w_n$ take the form
\begin{eqnarray}
 \label{eq:pn}
 p_n &=(x^2+y^2-1)^{\frac n 2}P_n\left(\frac{xy}{\sqrt{x^2+y^2-1}}\right),\\
 \label{eq:wm}
 w_m &=-\frac{\mathrm{sgn}(\xi_m)}{\sqrt{(xy-\xi_m)^2-(1-x^2)(1-y^2)}}.
\end{eqnarray}
In particular --- recall that $\pm1$ are necessarily among the zeros $\xi_m$ --- we have
\begin{equation}
 w_m=\frac{1}{x\mp y}
 \quad\mbox{for}\quad
 \xi_m=\pm1.
\end{equation}

The above formulae allow us to express the integrals in \eqref{eq:inteq_decomp}, \eqref{eq:constraint_decomp} and \eqref{eq:Ernst_decomp} in terms of the functions $p_n$ and $w_m$. The result takes a particularly elegant form if we define the following linear combinations of $p_n$ and $w_m$,
\begin{equation}\label{eq:defST}
 S_n = \sum_{k=0}^{N-n}\alpha_{k+n}p_k,
 \quad
 T_{mn} = \sum_{k=0}^{N-n-1}\beta_{m,k+n}p_k-\bar\beta_{m,n-1}w_m
\end{equation}
for $m=1,\dots,N$, $n=0,\dots,N$. For $n=0$, the second term in $T_{m0}$ contains\footnote{Here it is useful to extend the previous definitions of $c_{mk}$ and $d_{mk}$ to also include the case $k=-1$.}
\begin{equation}
\fl
 -\bar\beta_{m,-1}=-c_{m,-1}+\ii e_{m,-1}
 = -\sum_{k=0}^Nc_k\xi_m^k+\ii\sum_{k=0}^Nd_k\xi_m^k
 =-\bar\E\p(\xi_m)
 =\ii b\p(\xi_m),
\end{equation}
i.e.\ a term that also appears in \eqref{eq:Ernst_decomp}. This allows us to write the formula for the Ernst potential in the same form as the terms in the integral equation, and we obtain the following algebraic equations,
\begin{eqnarray}
 \label{eq:EQ1}
 \mbox{integral equation:} \quad
  & S_{l+1}A_0+\sum_{m=1}^NT_{m,l+1}A_m=0,\quad l=0,\dots,N-1,\\
  \label{eq:EQ2}
 \mbox{constraint:}
  & A_0+\sum_{m=1}^N w_m A_m = 1,\\
  \label{eq:EQ3}
  \mbox{Ernst formula:}
  & \E  = S_0A_0+\sum_{m=1}^{N}T_{m0} A_m.
\end{eqnarray}

\subsection{Calculation of the Ernst potential $\E$}\label{sec:EP}

With the above considerations, we arrive at a simple algorithm to algebraically construct the Ernst potential for an SGGTN solution with polynomial initial data:
\begin{enumerate}
 \item Select a polynomial degree $N$ and fix the initial Ernst potential \eqref{eq:initial_data} by choosing the zeros $\xi_m$, $m=1,\dots,N$, of $f\p$, together with a constant $c$, and the polynomial coefficients $d_k$, $k=0,\dots,N$, of $b\p$ such that the constraints described below \eqref{eq:initial_data} are satisfied. Also calculate the coefficients $c_k$, $k=0,\dots,N$, in the power form \eqref{eq:ck} of $f\p$, and the complex constants $\alpha_k=c_k+\ii d_k$, cf.\ \eqref{eq:alphabeta}.
 
 \item Calculate the constants $c_{mk}$ and $e_{mk}$, $m=1,\dots,N$, $k=-1,\dots, N-1$, from \eqref{eq:cmk1} and \eqref{eq:eml} 
 (with the above discussed extended range for the second index including $k=-1$) as well as their complex combinations $\beta_{mk}=c_{mk}+\ii e_{mk}$, cf.\ \eqref{eq:alphabeta}.
 
 \item Calculate the functions $S_n$ and $T_{mn}$ for $m=1,\dots,N$, $n=0,\dots,N$, from \eqref{eq:defST} in terms of $p_n$ and $w_m$, defined in 
 \eqref{eq:pn} and \eqref{eq:wm}, respectively.
 
 \item Solve the $N+1$ linear equations \eqref{eq:EQ1}, \eqref{eq:EQ2} for the $N+1$ unknowns $A_0$, $\dots$, $A_N$.
 
 \item Plug the result into \eqref{eq:EQ3} to obtain the Ernst potential $\E$ in the entire Gowdy square $(x,y)\in[-1,1]^2$.
\end{enumerate}
This algorithm can easily be implemented in computer algebra software like Maple or Mathematica, in order to obtain explicit expressions, and for purposes of symbolic calculations with the resulting Ernst potential.
\JH{For ease of implementation, we provide a pseudocode formulation in \ref{Sec:AppPseudocode}.}
Alternatively, one can use numerical packages like Matlab, if one is only interested in highly accurate numerical values of the Ernst potential.

In the above algorithm, we first work out the unknown functions $A_0$, $\dots$, $A_N$, and then use \eqref{eq:EQ3} to calculate the Ernst potential.
Alternatively, we can also construct an explicit formula from which we obtain $\E$ directly. For that purpose, we consider $\E$ to be an additional unknown and take \eqref{eq:EQ3} as an additional equation. Then \eqref{eq:EQ1}--\eqref{eq:EQ3} is a system of $N+2$ equations for the $N+2$ unknowns $\E$, $A_0$, \dots $A_N$. Applying Cramer's rule, we find an exact expression for $\E$, which simplifies to
\begin{equation}\label{eq:Edet}
 \E = \frac{ 
      \left|\begin{array}{cccc}
     S_0    & S_1    & \dots & S_N\\
     T_{10} & T_{11} & \dots & T_{1N}\\
     \vdots & \vdots &       & \vdots\\
     T_{N0} & T_{N1} & \dots & T_{NN}
    \end{array}\right|}
    {\left|\begin{array}{cccc}
    1      & S_1    & \dots & S_N\\
    w_1    & T_{11} & \dots & T_{1N}\\
    \vdots & \vdots &       & \vdots\\
    w_N    & T_{N1} & \dots & T_{NN}
    \end{array}\right|}
    =:\frac{A}{B}.
\end{equation}

Coming back to the algorithm,
a minor simplification in the calculation of $\E$ is achieved by observing that a simple expression can be found for one of the unknowns in the above system of equations, namely the function $A_0$. To this end, consider the integral equation \eqref{eq:EQ1} for $l=N-1$, which simplifies to
\begin{equation}\label{eq:alphaN}
 \alpha_N A_0 - \bar\alpha_N\sum_{m=1}^Nw_mA_m=0.
\end{equation}
Adding 
$\bar\alpha_N$ times the constraint \eqref{eq:EQ2} gives a very simple equation, which can be solved for $A_0$,
\begin{equation}\label{eq:A0}
 A_0 = \frac{\bar\alpha_N}{\alpha_N+\bar\alpha_N}
\equiv \frac{c_N-\ii d_N}{2c_N}
=\mathrm{constant}.
\end{equation}
Hence, it remains to solve the $N$ equations \eqref{eq:EQ1} for $A_1$, $\dots$, $A_N$.

This result can also be used to simplify the calculation of the $(N+1)\times(N+1)$ determinant $B$ in \eqref{eq:Edet}: Since the formula for $A_0$ can alternatively be derived from Cramer's rule, we can equate the resulting expression (which also contains $B$) with \eqref{eq:A0}  and solve for $B$. In this way, we obtain the simplified formula
\begin{equation}
 B=
   \frac{2c_N}{c_N-\ii d_N}
    \left|\begin{array}{ccc}
           T_{11} & \dots & T_{1N}\\
           \vdots &       & \vdots\\
           T_{N1} & \dots & T_{NN}
          \end{array}\right|,
\end{equation}
which only requires an $N\times N$ determinant.

\subsection{Calculation of the metric potentials $u$ and $Q$}

Once we have constructed the Ernst potential $\E$, we can obtain the metric potentials $u$ and $Q$ from \eqref{eq:uQ}. 
To this end, we calculate the real part $f$ of $\E$ and the auxiliary quantity $a$, which are required in \eqref{eq:uQ}. 

From \eqref{eq:Edet}, we simply have
\begin{equation}\label{eq:ff}
 f=\Re(\E)=\frac{A\bar B+\bar A B}{2B\bar B}.
\end{equation}
Next, in order to calculate $a$, we evaluate the integral representation \eqref{eq:a2}. Using \eqref{eq:mu}, we find
\begin{equation}
 \xi\mu(\xi)
 = \xi A_0+\sum_{m=1}^N\left(1+\frac{\xi_m}{\xi-\xi_m}\right) A_m.
\end{equation}
With the previously discussed integrals, we then have
\begin{equation}\label{eq:atilde}
 \hat a
 :=  \frac{1}{\pi}\int_{-1}^1\frac{\xi\mu(\xi)}{\sqrt{1-\sigma^2}}\,\dd\sigma
 = \zeta A_0+\sum_{m=1}^N (1+\xi_m w_m) A_m.
\end{equation}
Therefore, one approach for calculating $a$ is to apply the algorithm in Sec.~\ref{sec:EP} to obtain $A_0$, $\dots$, $A_N$ as well as the resulting Ernst potential and its real part $f$, and then to use the formula
\begin{equation}\label{eq:aFormula}
 a = \frac{2}{f}\Im(\hat a)
   = \frac{2}{f}\Im\left(\zeta A_0+\sum_{m=1}^N (1+\xi_m w_m) A_m\right).
\end{equation}
Alternatively, we can consider $\hat a$ to be an additional unknown and solve \eqref{eq:EQ1}, \eqref{eq:EQ2}, \eqref{eq:atilde} for $\hat a$ with Cramer's rule. Since the equation \eqref{eq:atilde} for $\hat a$ can formally be obtained from the equation \eqref{eq:EQ3} for $\E$ with the replacements $S_0\to\zeta$, $T_{m0} \to 1+\xi_m w_m$, we find a formula for $\hat a$ with the same replacements in the determinant representation \eqref{eq:Edet} for $\E$. The result is
\begin{equation}
 \hat a = \frac{C}{B},\quad
 C:=\left|\begin{array}{cccc}
           \zeta       & S_1    & \dots & S_N\\
           1+\xi_1 w_1 & T_{11} & \dots & T_{1N}\\
           \vdots      & \vdots &       & \vdots\\
           1+\xi_N w_N & T_{N1} & \dots & T_{NN},
          \end{array}\right|.
\end{equation}
Hence, using \eqref{eq:ff}, we finally obtain
\begin{equation}
 a = -2\ii \frac{C\bar B - \bar C B}{A\bar B + \bar A B}.
\end{equation}

These expressions for $f$ and $a$ then allow us to calculate $u$ and $Q$ from \eqref{eq:uQ}.

\section{Example 1: Initial data of fourth degree with a double root\label{sec:Ex1}}

We illustrate the construction of SGGTN solutions with polynomial initial data with two examples. Firstly, we show that the above procedure can also handle initial functions $f\p$ with multiple roots through a suitable limiting process. Alternatively, one could also extend the above algorithm to allow roots $\xi_i$ of arbitrary multiplicities from the outset. This requires a generalisation of the formula \eqref{eq:mu} for $\mu$, see \cite{MankoSibgatullin1993}. However, in favour of simple equations and a concise determinant representation of the solutions, we have assumed single roots. Nevertheless, we can simply start from distinct roots and then consider the limit where some roots approach each other.

As an example, we \JH{choose $N=4$ and} consider an initial function $f\p$ with the four initially distinct roots
\begin{equation}
 \xi_1 = 1,\quad
 \xi_2 = -1,\quad
 \xi_3 = 2,\quad
 \xi_4 = 2 + \eps.
\end{equation}
Later, we will perform the limit $\eps\to 0$, in which $f\p$ will become a polynomial of fourth degree with a double root at $2$. We also take $c=-1$ in \eqref{eq:initial_data}, so that
\begin{equation}
 f\p(\zeta) = -(\zeta^2 - 1)(\zeta - 2)(\zeta - 2 - \eps)
\end{equation}
is positive in $(-1,1)$ as required. Finally, we choose the imaginary part
\begin{equation}
 b\p(\zeta)=-\zeta^2,
\end{equation}
which satisfies the last condition in \eqref{eq:Ernstconstraints}.

\begin{figure}\centering
 \includegraphics[width=0.45\linewidth]{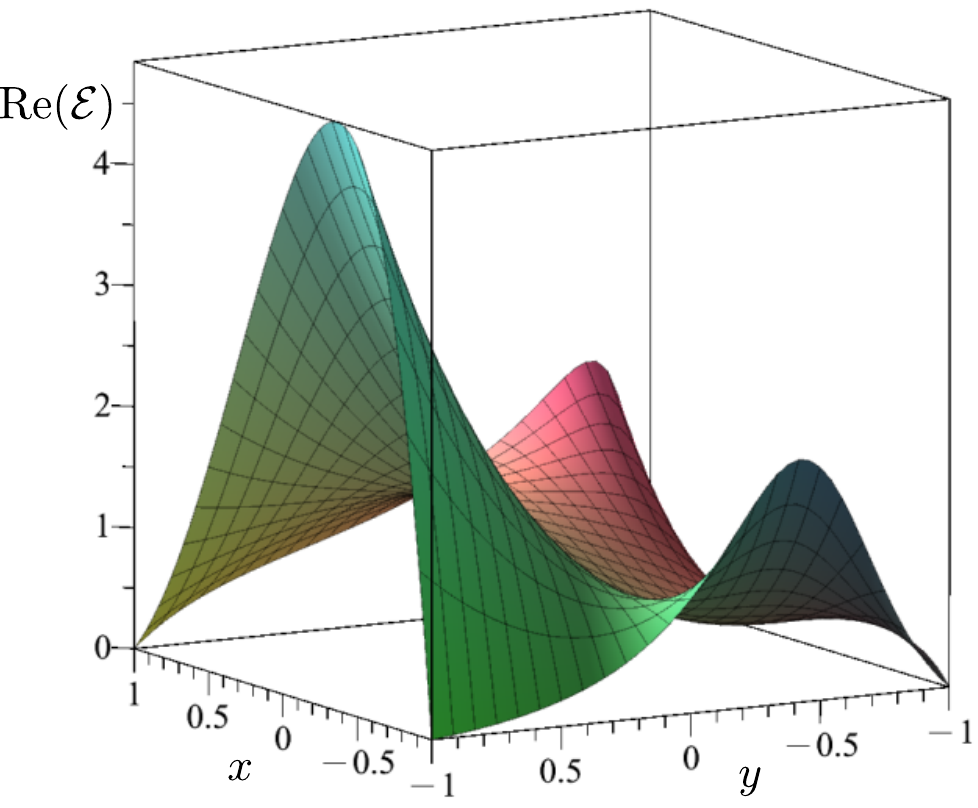}
 \quad
 \includegraphics[width=0.45\linewidth]{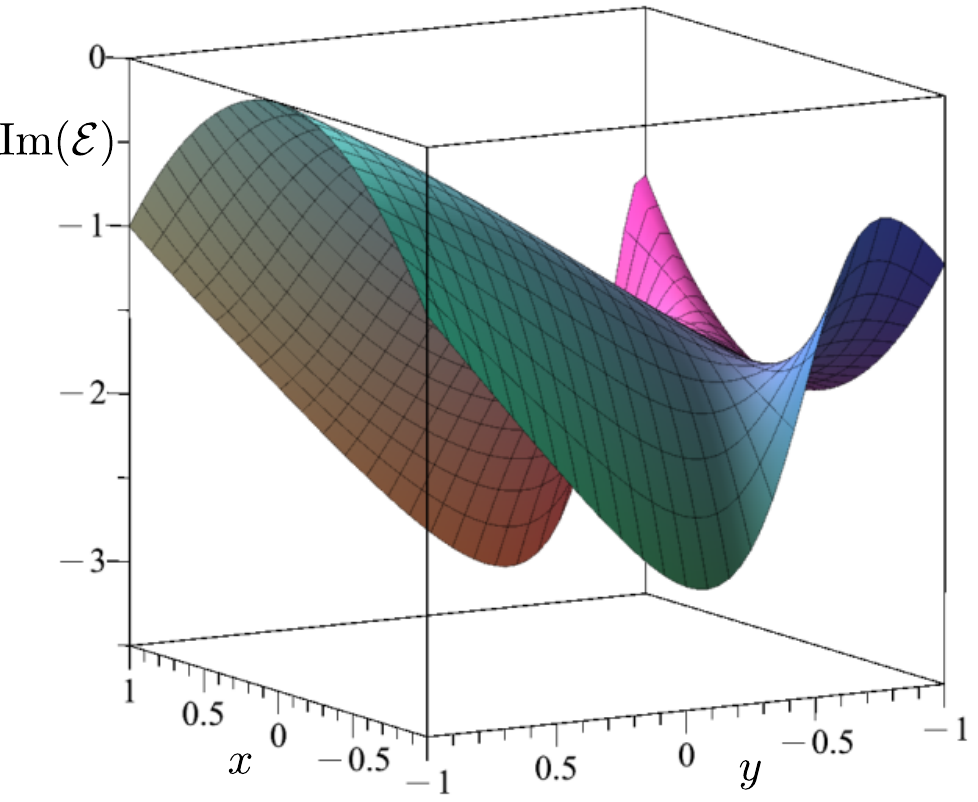}
 \caption{The Ernst potential \eqref{eq:EP1} for the solution with initial data of degree four (Example 1).}
 \label{fig:EP1}
\end{figure}

Following the algorithm from Sec.~\ref{sec:EP}, we first obtain the following  polynomial coefficients $\alpha_k=c_k+\ii d_k$ \JH{in the representation  $\E\p(\zeta)\equiv f\p(\zeta)+\ii b\p(\zeta)=\sum_{k=0}^N\alpha_k\zeta^k$},
\begin{equation}
\fl
 \alpha_0 = 4 + 2\eps,\quad
 \alpha_1 = -4 - \eps,\quad
 \alpha_2 = -3 - 2\eps - \ii,\quad
 \alpha_3 = 4 + \eps,\quad
 \alpha_4 = -1.
\end{equation}
Next, we calculate the constants $\beta_{mk}=c_{mk}+\ii e_{mk}$ \JH{from \eqref{eq:cmk1} and \eqref{eq:eml}},
\begin{eqnarray}
  \fl\nonumber
  \beta_{1\, -1} = -\ii,\quad
  \beta_{10} = -4 - 2 \eps - \ii,\quad
  \beta_{11} = -\eps - \ii,\quad
  \beta_{12} = 3 + \eps,\quad
  \beta_{13} = -1,\\ 
  \fl\nonumber
  \beta_{2\, -1} = -\ii,\quad
  \beta_{20} = 4  + 2\eps +\ii,\quad
  \beta_{21} = -8 - 3\eps - \ii,\quad
  \beta_{22} = 5 + \eps,\quad
  \beta_{23} = -1,\\ 
  \fl\nonumber
  \beta_{3\, -1} = -4\ii,\quad
  \beta_{30} =  -2 - \eps - 2\ii,\quad
  \beta_{31} =  1 - \ii,\quad
  \beta_{32} = 2 + \eps,\quad
  \beta_{33} = -1,\\ 
  \fl
  \beta_{4\, -1} = -(2 + \eps)^2\ii,\quad
  \beta_{40} = - 2 - (2+ \eps)\ii,\quad
  \beta_{41} = 1 - \ii,\quad
  \beta_{42} = 2,\quad
  \beta_{43} = -1.
\end{eqnarray}
We also need the polynomials $p_n$ up to $n=4$, which read in terms of $x$ and $y$ \JH{[cf.~\eqref{eq:pn}]}
\begin{eqnarray}
 \nonumber
 p_0 = 1,\quad
 p_1 = xy,\quad
 p_2 = \frac12\left[(3y^2 - 1)x^2 - y^2 + 1\right],\\
 \nonumber
 p_3 = \frac12 xy\left[(5y^2 - 3)x^2 - 3(y^2 - 1)\right],\\
p_4 = \frac18\left[(35y^4 - 30y^2 + 3)x^4 - 2(15y^4 - 18y^2 + 3)x^2 
        + 3(y^2-1)^2\right].
\end{eqnarray}
Continuing with the algorithm, \JH{we next calculate the functions $S_n$ and $T_{mn}$ from \eqref{eq:defST} and plug the results into the integral equation in form of the algebraic equations \eqref{eq:EQ1}. The function $A_0$ can directly be obtained from the explicit formula
$A_0=(c_N-\ii d_N)/(2c_N)=1/2$. Hence we do not need to consider the constraint \eqref{eq:EQ2} and can obtain the remaining functions $A_1$, $\dots$, $A_4$ by solving the four equations \eqref{eq:EQ1}. Plugging the solution into \eqref{eq:EQ3}, we}
can now easily calculate the Ernst potential $\E$, and then perform the limit $\eps\to0$. The result is
\begin{equation}
\label{eq:EP1}
 \E(x,y) = \frac{q_1 r + q_2 + \ii(q_3 r + q_4)}
                {q_5 r + q_6 + \ii(q_7 r + q_8)}
\end{equation}
with the square root function
\begin{equation}
 r = \sqrt{x^2 - 4xy + y^2 + 3}
\end{equation}
and the real polynomials
\begin{eqnarray}
\fl\nonumber 
 q_1 &=& (1-y^2)^3 [2x^8 + 4(5 - y)x^7 + (93 - 25y)x^6 + 4(65 - 9y)x^5]  \\
\fl\nonumber
	 &&\quad + (25y^7 - 65y^6 + 681y^5 + 1455y^4 - 285y^3 - 1499y^2 + 155y + 557)x^4 \\
\fl\nonumber
	 &&\quad - (20y^7 - 132y^6 - 60y^5 + 1548y^4 + 2492y^3 + 1012y^2 - 916y - 380)x^3  \\
\fl\nonumber
	 &&\quad - (45y^7 - 145y^6 + 513y^5 + 1515y^4 - 329y^3 - 2275y^2 - 613y + 777)x^2  \\
\fl\nonumber
	 &&\quad - (20y^7 - 148y^6 - 60y^5 - 708y^4 - 2372y^3 - 2364y^2 + 1132y + 916)x \\
\fl  &&\quad - 5y^7 + 15y^6 - 93y^5 - 225y^4 - 119y^3 + 149y^2 - 1063y - 195,\\
\fl\nonumber
 q_2 &=& (1-y^2)^3 [2x^9 + 4(5 - 2y)x^8 + (5y^2 - 65y + 96)x^7 + 2(10y^2 - 111y + 145)x^6]\\
\fl\nonumber 
     &&\quad - (45y^8 - 185y^7 + 857y^6 + 1311y^5 - 1581y^4 - 915y^3 + 2075y^2 + 365y - 692)x^5\\
\fl\nonumber	
	&&\quad - (20y^8 - 110y^7 - 162y^6 - 2190y^5 - 4602y^4 - 1946y^3 + 1802y^2 - 234y - 770)x^4\\
\fl\nonumber	
	&&\quad + (25y^8 - 165y^7 + 445y^6 + 567y^5 - 4565y^4 - 8447y^3 - 5233y^2 - 147y - 144)x^3\\
\fl\nonumber	
	&&\quad + (36y^8 - 310y^7 + 94y^6 - 1230y^5 - 4354y^4 - 66y^3 + 8410y^2 + 4934y - 346)x^2\\
\fl\nonumber	
	&&\quad + (25y^8 - 85y^7 + 495y^6 + 939y^5 + 2705y^4 + 6697y^3 + 8941y^2 + 385y - 1158)x\\
\fl	
	&&\quad + 4y^8 - 30y^7 - 6y^6 - 270y^5 - 1118y^4 - 3082y^3 - 4162y^2 - 5322y - 1374,\\
\fl\nonumber	
q_3 &=& 4(1-y^2)[2(11y^2 + 1)x^6 - 4(11y^3 - 13y^2 + 25y + 1)x^5 - (65y^3 - 141y^2 + 347y\\
\fl\nonumber
	&&\quad - 39)x^4 + 8(9y^3 + 23y^2 - 63y + 35)x^3] 
	- (440y^5 + 224y^4 - 1360y^3 + 1856y^2
	\\
\fl\nonumber	
	&&\quad   + 344y - 2528)x^2
	+ 16(1 + y)(7y^4 + 20y^3 - 2y^2 - 148y + 139)x\\
\fl	
	&&\quad + 180y^5 + 236y^4 - 120y^3 - 1256y^2 - 380y + 828,\\
\fl\nonumber
q_4 &=& 4(1-y^2)[2(11y^2 + 1)x^7 - 4(22y^3 - 13y^2 + 26y + 1)x^6 + (55y^4 - 169y^3 + 371y^2\\
\fl\nonumber
	&&\quad - 339y + 42)x^5 + 2(26y^4 - 33y^3 + 481y^2 - 363y + 137)x^4]\\
\fl\nonumber	
	&&\quad + (360y^6 + 72y^5 - 2456y^4 + 5488y^3 - 1064y^2 - 6136y + 2456)x^3\\
\fl\nonumber	
	&&\quad + (352y^6 - 16y^5 - 128y^4 + 4576y^3 - 3872y^2 - 2896y + 5056)x^2\\
\fl\nonumber
	&&\quad - (140y^6 + 108y^5 - 416y^4 - 2088y^3 + 3980y^2 + 2812y - 3000)x\\
\fl
	&&\quad - 144y^6 - 88y^5 - 120y^4 + 16y^3 - 1280y^2 + 456y + 2184,\\
\fl\nonumber	
q_5 &=& 16(1 + y)\bigg[(1 - y)[-2x^4 + 4(y - 3)x^3 + 3(5y - 11)x^2 + 4(3y - 13)x]
\\
\fl
	&&\quad - 5y^2 + 14y - 45\bigg],\\
\fl\nonumber
q_6 &=& 16(1 + y)\bigg[(1 - y)[-2x^5 + 4(2y - 3)x^4 - (5y^2 - 39y + 36)x^3 - 2(6y^2 - 39y + 35)x^2]\\
\fl
 &&\quad + (3y - 2)(5y^2 - 14y + 45)x + 2(2y^3 - 11y^2 + 12y - 39)\bigg],\\
\fl\nonumber
q_7 &=& 64(2x^2 + 4(1 - y)x - 5y + 3),\\
\fl
q_8 &=& 64(2x^3 - 8x^2y + 5xy^2 + 4x^2 - 13xy + 4y^2 + 6x - 6y + 6).
\end{eqnarray}
Corresponding plots of the real and imaginary parts of $\E$ can be found in Fig.~\ref{fig:EP1}.

\begin{figure}\centering
 \includegraphics[width=0.45\linewidth]{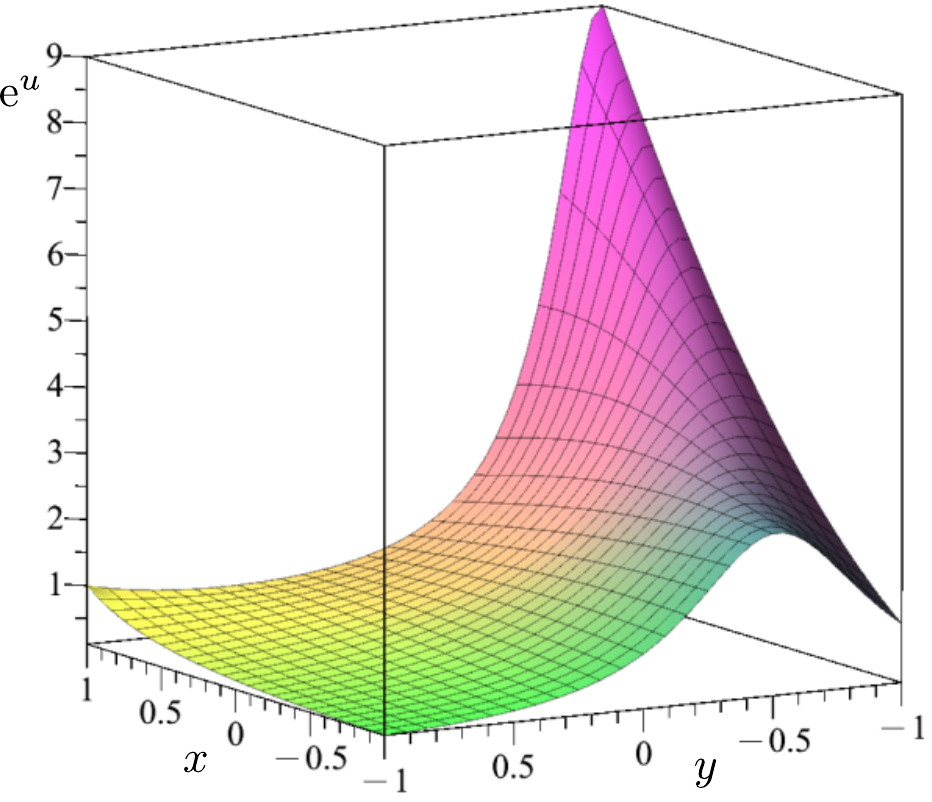}
 \quad 
 \includegraphics[width=0.45\linewidth]{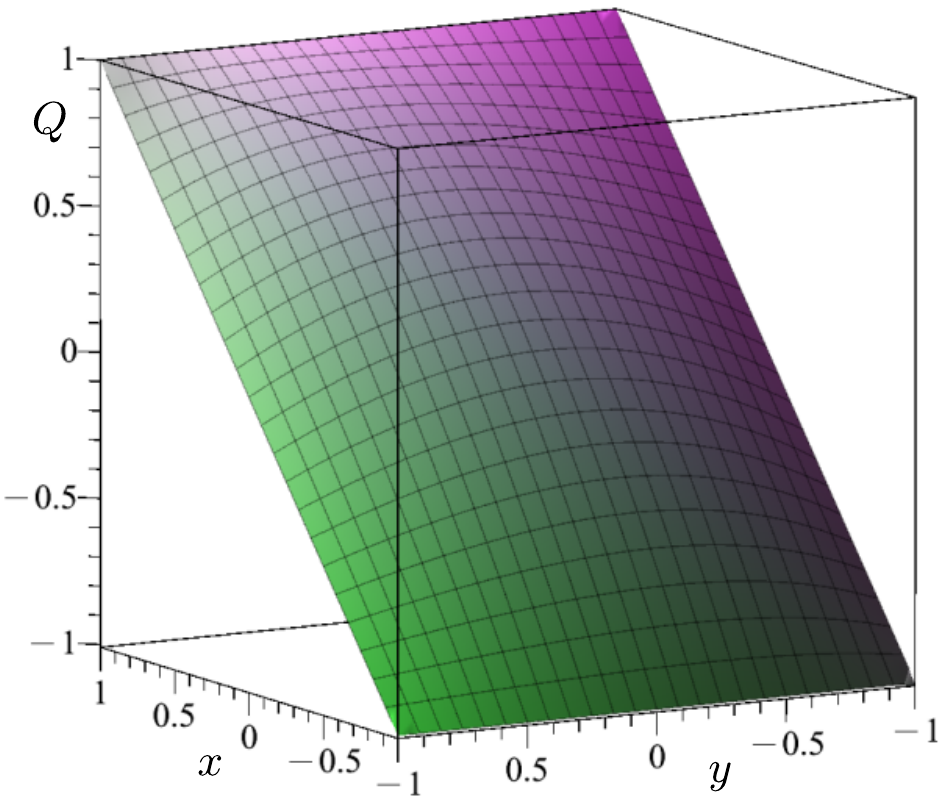}
 \caption{The metric potentials $\ee^u$ and $Q$ for Example 1.}
 \label{fig:uQ1}
\end{figure}

Next, we calculate the quantity $a$ \JH{from \eqref{eq:aFormula}}. The result is
\begin{equation}
 a=\frac{32}{f}\,\Im\left(\frac{\tilde q_1 r + \tilde q_2 + \ii(\tilde q_3 r + \tilde q_4)}{q_5r + q_6 + \ii(q_7 r + q_8)}\right)
\end{equation}
with the same functions $q_5,\dots,q_8$ and $r$ as in \eqref{eq:EP1}, and with
\begin{eqnarray}
\fl\nonumber
 \tilde q_1 &=& (1-y^2)[2yx^5 - 4(1 - y)^2x^4 - (10y^2 - 17y + 21)x^3 + 16(y - 3)x^2 + 4(y - 3)(1 + y)]\\
 \fl
 &&\quad- (1 + y)(10y^3 - 9y^2 - 24y + 59)x,\\
\fl\nonumber
\tilde q_2 &=& (1-y^2)[2yx^6 - 4(2y^2 - 2y + 1)x^5 + (5y^3 - 26y^2 + 28y - 21)x^4 \\
\fl\nonumber
&&\quad + 2(4y^3 - 14y^2+ 35y - 27)x^3 - (5y^2 - 6y + 21)(1 + y)]\\
\fl
&&\quad- (1 + y)[5y^3 + 56y^2 - 183y + 86)x^2 - 2(4y^4 - 14y^3 - 3y^2 + 28y - 51)x],\\
\fl
\tilde q_3 &=& 8yx^3 - 16(y^2 + 1)x^2 + x(12y - 20) + 16y(1 + y),\\
\fl\nonumber
\tilde q_4 &=& 8yx^4 - 16(2y^2 + 1)x^3 + 4(5y^3 + 14y - 5)x^2 + 8(2y^2 + 7y - 3)x\\
\fl
&&\quad- 4(1 + y)(5y^2 + 3).
\end{eqnarray}
From $a$ and $f=\Re(\E)$, we can finally calculate the metric functions $\ee^u$ and $Q$ \JH{using \eqref{eq:uQ}}. While exact expressions can easily be obtained, they are somewhat lengthy. Hence, we restrict ourselves to plotting these functions in Fig.~\ref{fig:uQ1},
and we explicitly give only the significantly simpler function values on the boundaries of the Gowdy square,
\begin{equation}
 \ee^u=\left\{ 
        \begin{array}{ll}
          \displaystyle
          \frac{4(y - 2)^2}{y^6 - 6y^5 + 9y^4 + 8y^3 - 23y^2 - 2y + 17},
          & x=1,\\[2ex]
          \displaystyle
          \frac{4(y + 2)^2}{y^6 + 10y^5 + 41y^4 + 88y^3 + 105y^2 + 62y + 17},
          & x=-1,\\[2ex]
          \displaystyle
          \frac{1}{(x - 2)^2},
          & y=1,\\[2ex]
          (x + 2)^2,
          & y=-1
        \end{array}
        \right.
\end{equation}
\begin{equation}
 Q = \left\{ 
	  \begin{array}{ll}
	   \pm1, & x=\pm1,\\
	   x,    & y=\pm1.
	  \end{array}
	  \right.
\end{equation}

\section{Example 2: Initial data of sixth degree\label{sec:Ex2}}

Next, we want to construct an interesting solution with a singularity. For that purpose, we choose suitable initial data of sixth degree. Specifically, we consider the initial Ernst potential $\E\p=f\p+\ii b\p$ with
\begin{equation}\label{eq:ID2}
 f\p(\zeta) = -(\zeta^2-1)(\zeta^2-4)(\zeta^2-9),\quad
 b\p(\zeta) = -\frac12 \zeta^5 - \zeta^2 + \frac52 \zeta.
\end{equation}
Here, $f\p$ is a polynomial of sixth degree with zeros at $\pm1$, $\pm2$, $\pm3$, and both $f\p$ and $b\p$ satisfy the conditions in \eqref{eq:Ernstconstraints}. Moreover, we have $b_B-b_A\equiv b\p(-1)-b\p(1)=-4$. Consequently, according to the remark below the regularity condition \eqref{eq:regularity}, the resulting Ernst potential will be singular at $\theta=t=\pi$, corresponding to $x=y=-1$.

Following again the algorithm from Sec.~\ref{sec:EP}, we obtain the Ernst potential for all $x$, $y$ in the Gowdy square. Since the result is much more lengthy than the Ernst potential \eqref{eq:EP1} in our previous example\footnote{Maple's function \texttt{length}, which measures the size of an expression, returns a value of $2{,}700$ for the Ernst potential \eqref{eq:EP1} in Example 1 above. For the Ernst potential in Example 2, we obtain $86{,}357$.},
we do not give the full expression, but only provide boundary values and plots of $\E$.

\begin{figure}\centering
 \includegraphics[width=0.45\linewidth]{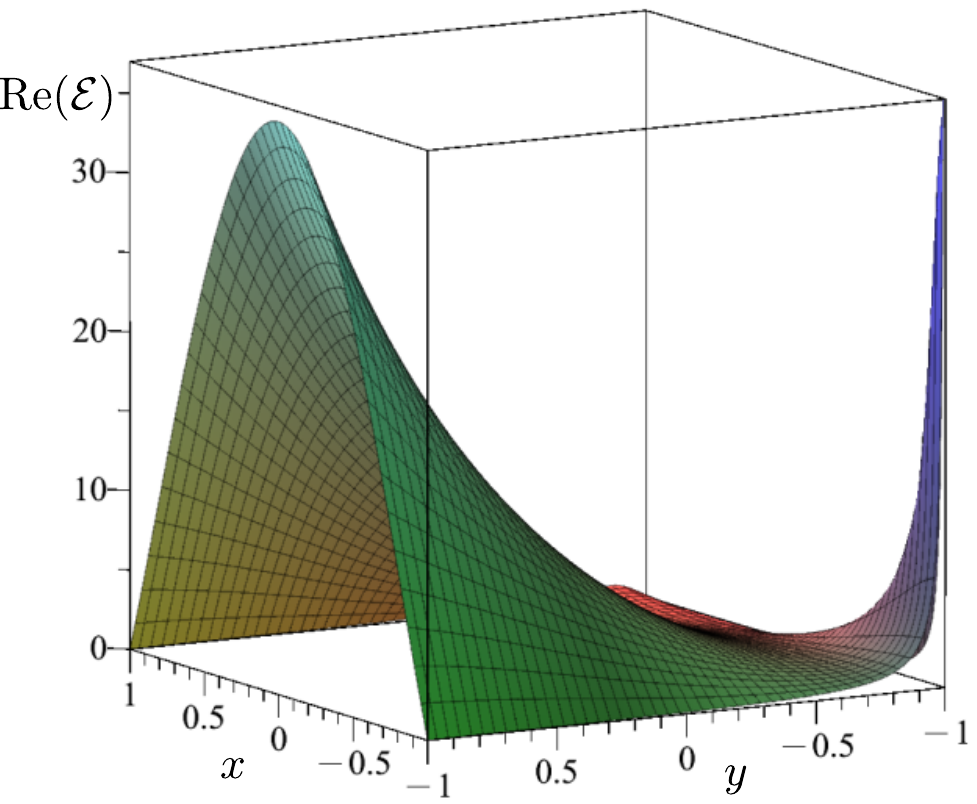}
 \quad
 \includegraphics[width=0.45\linewidth]{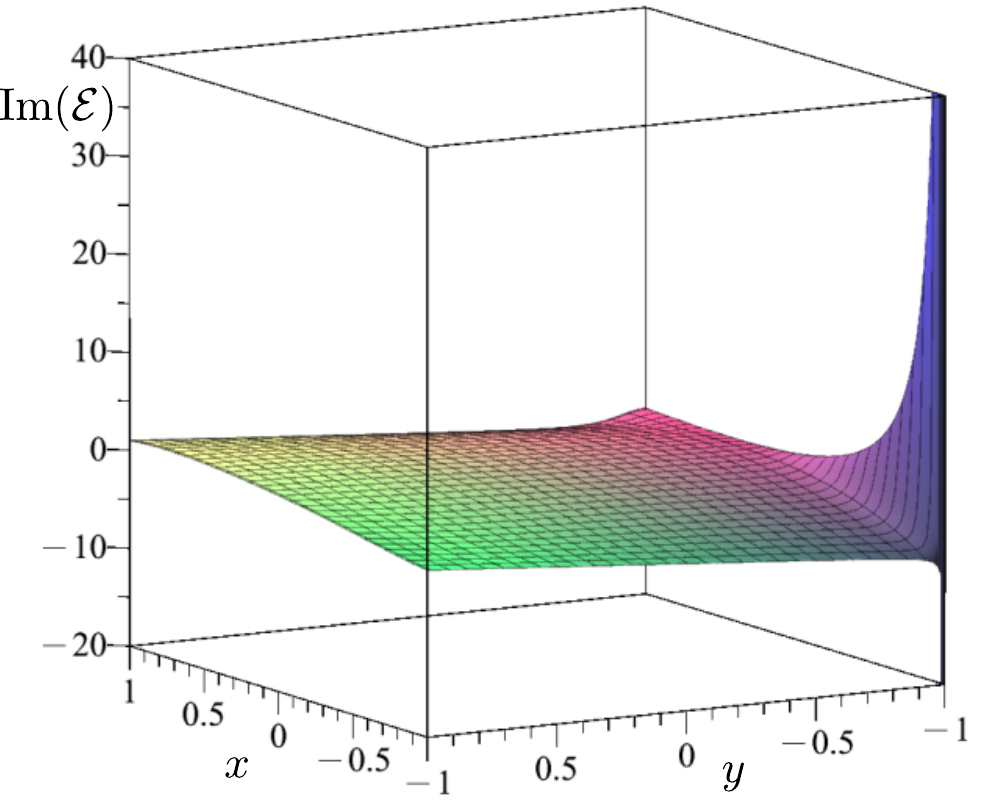}
 \caption{Ernst potential for the solution with initial data \eqref{eq:ID2} of degree six. The Ernst potential clearly diverges at $x=y=-1$.}
 \label{fig:EP2}
\end{figure}

Besides the initial data surface $y=1$, where $\E$ is given by \eqref{eq:ID2}, we find the following values at the other three boundaries of the Gowdy square,
\begin{equation}
\fl
 \E=\left\{ 
    \begin{array}{ll}
     \displaystyle
     \frac{(1 - y)(2y^4 + 3y^3 + 4y^2 + 9y + 2) + 2\ii(1 + y)(y^2 - 4)(y^2 - 9)(2y - 1)}{(y + 2)\left[2(1 + y)(y - 2)(y^2 - 9) - \ii(1 - y)(y^2 + 3)\right]}, 
      & x=1,\\[2ex]
     \displaystyle
     \frac{(y - 1)(2y^4 - y^3 - 4y^2 - 11y - 18) + 2\ii(1 + y)(y^2 - 4)(y^2 - 9)(2y - 5)}{(1 + y)\left[2(y^2 - 4)(y^2 - 9) + \ii(1 - y)(y^2 + y + 2)\right]},
      & x=-1,\\[2ex]
      \displaystyle
      \frac{2(x - 2)\left[2(x + 2)(x^2 - 9)(2x^2 - 3x - 1) + \ii(1-x^2)(2x^2 + x + 5)\right]}{2(1 + x)\left[(1 - x)(x^2 + x + 2) - 2\ii(x^2 - 4)(x^2 - 9)\right]},
      & y=-1.
    \end{array}\right.
\end{equation}
It is already obvious from these expressions that $\E$ diverges at $x=y=-1$. This is also clearly visible in the plots shown in Fig.~\ref{fig:EP2}.

\begin{figure}\centering
 \includegraphics[width=0.45\linewidth]{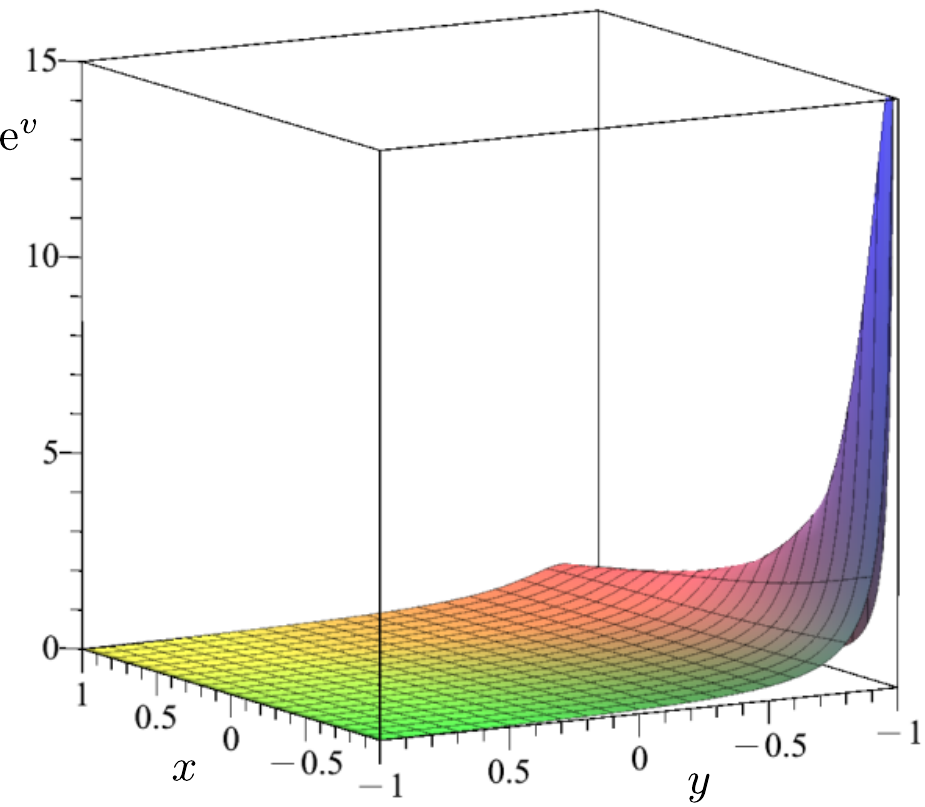}
 \quad
 \includegraphics[width=0.45\linewidth]{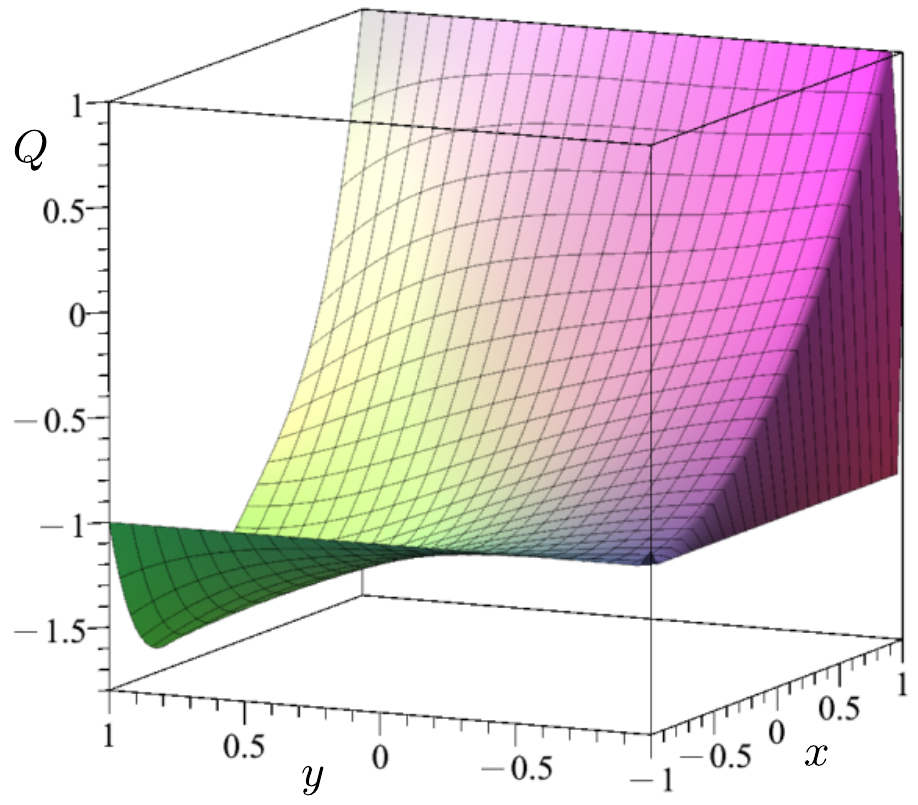}
 \caption{The metric potentials $\ee^v\equiv (1-y^2)\ee^u$ and $Q$ for the solution with initial data of degree six.}
 \label{fig:vQ2}
\end{figure}

Continuing with the algorithm, we obtain $a$ and the resulting metric potentials $\ee^u$ and $Q$. 
In this example, $\ee^u$ diverges at the future boundary $y=-1$. This, however, is not related to the curvature singularity at one point of this boundary, but a general feature of many solutions in this parametrisation of the metric. In order to obtain a regular function, we can easily replace $\ee^u$ by $\ee^v:= (1-y^2)\ee^u\equiv \frac{1}{R_0}g(\xi,\xi)$, which is regular except at the curvature singularity.
Again, we restrict ourselves to plotting these functions, see Fig.~\ref{fig:vQ2}, 
and we give the reasonably simple boundary values,
\begin{equation}\label{eq:u}
  \fl
 \ee^u = \left\{
   \begin{array}{ll}
     \scalebox{0.9}
    {$\displaystyle\frac{16(y - 2)(y^2 - 9)}{(y + 2)(4y^8 - 8y^7 - 83y^6 + 158y^5 + 563y^4 - 948y^3 - 1245y^2 + 1278y + 1305)}$},
    & x=1,\\[2ex]
    \displaystyle
    \frac{16(y^2 - 4)(y^2 - 9)}{(1 + y)^2(4y^8 - 103y^6 + 966y^4 - 4y^3 - 3743y^2 - 4y + 5188)},
    & x=-1,\\[2ex]
    \displaystyle
    \frac{1}{(x^2 - 4)(x^2 - 9)},
    & y=1,
   \end{array}\right.
\end{equation}
\begin{equation}\label{eq:v}
\fl
 \ee^v = \frac{16(1 - x)(x^2 - 4)(x^2 - 9)}{(1 + x)(4x^8 - 103x^6 + 966x^4 - 4x^3 - 3743x^2 - 4x + 5188)}
 \quad \mbox{at} \quad y=-1,
\end{equation}
\begin{equation}
\fl
 Q = \left\{
   \begin{array}{ll}
    \pm1, & x=\pm1,\\[2ex]
    \displaystyle
    \frac54 x^4 + x - \frac54, & y=1,\\[2ex]
    1, & y=-1.
   \end{array}\right.
\end{equation}
Eq.~\eqref{eq:v} and the second equation in \eqref{eq:u} clearly show that $\ee^v$ diverges at the singularity at $x=y=-1$. Otherwise, $\ee^v$ is finite.

\section{Discussion}

We have described a simple algorithm that allows us to construct the Ernst potential $\E$, the auxiliary quantity $a$ and the metric potentials $u$ and $Q$ for smooth Gowdy-symmetric generalised Taub--NUT (SGGTN) solutions with polynomial initial Ernst potential at the past Cauchy horizon. All required integrals have been obtained, and the remaining calculations are purely algebraic. Alternatively, explicit formulae for $\E$ and $a$ in terms of determinants have been provided. 

As a practical application of the method, two SGGTN solutions have been constructed with initial data of fourth and sixth degrees, respectively. The former solution is regular everywhere inside the Gowdy square and at the future Cauchy horizon, while the latter solution was designed to develop a singularity.

The algorithm can easily be used for both symbolic and  numerical computer calculations. In this way, a large class of simple cosmological models becomes accessible, which can be studied in detail in the future.

\JH{A possible direction for future research would be to include electromagnetic fields in the present discussion. 
Since the coupled system of the Einstein--Maxwell equations in electrovacuum also belong to the class of integrable equations, it would be interesting to generalise our algorithm, in order to construct SGGTN \emph{electrovacuum} solutions with polynomial initial data. Additionally, simulating the dynamical evolution of small perturbations of these spacetimes could be particularly intriguing, as it would allow us to verify whether the Cauchy horizons disappear under sufficiently generic perturbations, in accordance with the strong cosmic censorship conjecture.}

\appendix
\section{Calculation of three integrals}\label{sec:AppIntegrals}

The three integrals \eqref{eq:int1}--\eqref{eq:int3} can be calculated as follows.

Firstly, to show that the principal value integral 
$\dashint_{-1}^1\frac{\dd\sigma}{\sqrt{1-\sigma^2}(\sigma-\tau)}$ vanishes, we can use the substitution $\sigma=\tanh z$ to obtain
$\dashint_{-\infty}^\infty\frac{\dd z}{\sinh z-\tau\cosh z}$, and calculate this with complex methods: Consider the integral of the same function over a closed contour from $-\infty$ to $z_0-\eps$ (where $z_0=\frac12\ln\frac{1+\tau}{1-\tau}$ is the zero of the denominator), followed by an upper semicircle to $z_0+\eps$, continuing along a straight line to $\infty$, then to $\infty+\ii\pi$, then a line to $z_0+\eps+\ii\pi$, a lower semicircle to $z_0-\eps+\ii\pi$, a line to $-\infty+\ii\pi$ and finally back to $-\infty$. The enclosed region is free of singularities, so the integral gives zero. On the other hand, it can be expressed as twice the required principal value integral plus terms of $\mathcal O(\eps)$. Hence, in the limit $\eps\to0^+$, we see that the principal value integral vanishes.

Alternatively, we can also evaluate the principal value integral directly, since an antiderivative can be found,
\begin{eqnarray}
\fl\nonumber
 \int \frac{\dd\sigma}{\sqrt{1-\sigma^2}(\sigma-\tau)}\\
 \fl
 \quad
 =-\frac{1}{\sqrt{1-\tau^2}}\ln\left[\frac{2}{|\sigma-\tau|}
 \left(\sqrt{(1-\sigma^2)(1-\tau^2)}+1-\sigma\tau\right)\right]=:g(\sigma).
\end{eqnarray}
Then it is easy to check that
\begin{equation}
 \lim_{\eps\to0^+}\left[g(\tau-\eps)-g(-1)+g(1)-g(\tau+\eps)\right]=0
\end{equation}
holds, i.e.\ the principal value integral vanishes.

Secondly, we calculate
$p_n(\rho,\zeta):=\frac{1}{\pi}\int_{-1}^1\frac{\xi^n}{\sqrt{1-\sigma^2}}\,\dd\sigma$. An explicit formula can be obtained by expanding $\xi^n \equiv (\zeta+\ii\rho\sigma)^n$. The resulting integrals $J_k:=\int_{-1}^1\frac{\sigma^k}{\sqrt{1-\sigma^2}}\,\dd\sigma$ can quickly be shown to satisfy $J_0=\pi$, $J_1=0$ and the recursion $J_{k+2}=\frac{k+1}{k+2}J_k$. Consequently, $J_k=0$ for odd $k$, and $J_{2k}=\frac{\pi(2k)!}{4^k(k!)^2}$ otherwise. From this, we obtain the power form
\begin{equation}
 p_n(\rho,\zeta) = n!\sum_{k=0}^{\lfloor \frac n2\rfloor}
			     \frac{(-1)^k\rho^{2k}\zeta^{n-2k}}{4^k(n-2k)!(k!)^2},
\end{equation}
where $\lfloor\dots\rfloor$ denotes the floor function.
This formula also shows that $p(\rho,\zeta)$ is even in $\rho$ and even/odd in $\zeta$ for even/odd $n$, respectively.

In order to obtain the alternative representation \eqref{eq:int2} in terms of Legendre polynomials, we can either start from the substitution $\sigma =\cos\phi$ and compare the resulting integral with the formula
$P_n(x)=\frac{1}{\pi}\int_0^\pi(x+\sqrt{x^2-1}\cos\phi)^n\,\dd\phi$ for Legendre polynomials, or we can derive a recursion formula for $p_n$ and relate it to the identity $(x^2-1)P'_n(x)=nxP_n(x)-nP_{n-1}(x)$.

Finally, we consider the integral
$w_m(\rho,\zeta):=\frac{1}{\pi}\int_{-1}^1\frac{\dd\sigma}{\sqrt{1-\sigma^2}(\xi-\xi_m)}$. With the same substitution $\sigma=\tanh z$ as above, we obtain
$w_m = \frac{1}{\pi}\int_{-\infty}^\infty
       \frac{\dd z}{\ii\rho\sinh z+(\zeta-\xi_m)\cosh z}$,
which we can calculate with an integration in the complex $z$-plane. We integrate along a contour $C$ with the four parts $C_1$: the real axis from $-\infty$ to $\infty$, $C_2$: the line from $\infty$ to $\infty+\ii\pi$, $C_3$: the line from $\infty+\ii\pi$ to $-\infty+\ii\pi$ and $C_4$: the line from $-\infty+\ii\pi$ to $-\infty$. The integrals over $C_2$ and $C_4$ make no contribution, and the other two integrals both equal $w_m$. Hence, $w_m$ is  $\ii\pi$ times the sum of residues in the enclosed strip $0\le\Im(z)\le\pi$ of the complex plane. The integrand turns out to have exactly one singularity in this region, namely at $z_0=\ii\arctan\frac{\rho}{\zeta-\xi_m}$. Evaluating the residue then gives the formula \eqref{eq:int3} for $w_m$.
\JH{
\section{Pseudocode}
\label{Sec:AppPseudocode}

A pseudocode formulation of the algorithm for calculating the Ernst potential, as described in Sec.~\ref{sec:EP}, is provided in Algorithm~\ref{Alg1} below. Our particular formulation is designed for systems that support symbolic variables and functions (e.g., Maple, Mathematica). For other systems, equivalent results can be obtained by implementing small additional subroutines.

\begin{algorithm}[H]
 \caption{Ernst potential $\E$ for an SGGTN solution with polynomial initial data}
 \label{Alg1}
 \algsetblock[Name]{Part}{Stop}{4}{6mm}
\JH{  
 \begin{algorithmic}[1]
  \algsetblock[Name]{Part}{Stop}{4}{6mm}
  \Part \textbf{: Initialisation}
  \State Define an integer $N$ 
	\Comment{polynomial degree}
  \State Define constants $\xi_m$, $m=1,\dots,N$ 
	\Comment{zeros of $f\p$ such that $\xi_1=1$, $\xi_2=-1$}
  \State Define a constant $c$ 
	\Comment{factor in $f\p$}
  \State Define constants $d_k$, $k=0,\dots,N$
	\Comment{polynomial coefficients of $b\p$}

  \algsetblock[Name]{Part}{Stop}{3}{6mm}	
  \Part \textbf{: Polynomial coefficients, partial fraction decompositions}
  \State  
   $f\p(\zeta):=c\prod_{m=1}^N(\zeta-\xi_m),\quad
   b\p(\zeta):=\sum_{k=0}^Nd_k\zeta^k$
   \Comment{symbolic functions}
  \For {$k$ from $0$ to $N$} 
  \State $c_k :=$ coefficient of $\zeta^k$ in $f\p$; \quad
         $d_k :=$ coefficient of $\zeta^k$ in $b\p$
  \State $\alpha_k := c_k + \ii d_k$
  \EndFor
  
  \For {$m$ from $1$ to $N$}
  \For {$k$ from $-1$ to $N - 1$}
   \State $c_{mk} := \sum_{l = 0}^{N-k-1} c_{k+l+1}\xi_{m}^l$;\quad
        $e_{mk} := \sum_{l = 0}^{N-k-1} d_{k+l+1}\xi_m^l$ 
   \State $\beta_{mk} := c_{mk} + \ii e_{mk}$;\quad 
        $\bar\beta_{mk} := c_{mk} - \ii e_{mk}$
  \EndFor
  \EndFor
  
  \algsetblock[Name]{Part}{Stop}{3}{6mm}
  \Part \textbf{: Functions $p$, $S$, $T$}
  \For {$n$ from $0$ to $N$}
  \State $p_n:= (\rho^2 + \zeta^2)^{n/2} \mathrm{LegendreP}(n, \zeta/\sqrt{\rho^2 + \zeta^2})$
   \Comment{symbolic functions $p_n$}
  \EndFor
  \For {$n$ from $0$ to $N$}
   \State $S_n := \sum_{k = 0}^{N - n}\alpha_{k + n}p_k$
   \Comment{symbolic functions $S_n$}
  \EndFor
  \For {$n$ from $0$ to $N$}
   \For {$m$ from $1$ to $N$}
   \State $T_{mn} := \sum_{k = 0}^{N-n-1}\beta_{m,k + n}p_k-\bar\beta_{m, n - 1}w_{m}$
    \Comment{symbolic functions $w_m$, $T_{mn}$}
   \EndFor
   \EndFor
   
  \algsetblock[Name]{Part}{Stop}{3}{6mm} 
  \Part \textbf{: Solve integral equation}
  \For {$l$ from $0$ to $N - 1$}
   \State $IntEQ_l := S_{l + 1} A_0 + \sum_{m = 1}^{N} T_{m, l + 1} A_m$
    \Comment{symbolic functions $IntEQ_l$}
  \EndFor
    \Comment{symbolic variables $A_m$}
  \State $A_0 := (c_N - \ii d_N)/(2 c_N)$
   \Comment{known solution for $A_0$}
  \State Solve $\{IntEQ_0,\dots,IntEQ_{N-1}\}$ for $\{A_1,\dots A_N\}$
   
  \algsetblock[Name]{Part}{Stop}{3}{6mm} 
  \Part\textbf{: Calculate Ernst potential} 
  \State $\E := S_0 A_0 + \sum_{m = 1}^{N} T_{m0} A_m$; \quad
  simplify and then substitute:
  \State 
   $\rho = \ii\sqrt{(1-x^2)(1-y^2)}$, \quad
   $\zeta = xy$, \quad
   $w_1 = 1/(x - y)$, \quad
   $w_2 = 1/(x + y)$
  \For {$m$ from $3$ to $N$}
   \State Substitute 
    $w_m = -\mathrm{sign}(\xi_m)/\sqrt{(xy - xmn)^2 - (1-x^2)(1-y^2)}$;
    \quad simplify $\E$ 
  \EndFor
  \State {\bf return} $\E$  
 \end{algorithmic} 
} 
\end{algorithm}

}



\end{document}